\documentclass[a4paper,10pt,twoside]{cpc-hepnp} 

\usepackage{multicol}
\usepackage{graphicx}
\usepackage{booktabs}
\usepackage{amssymb,bm,mathrsfs,bbm,amscd}
\usepackage[tbtags]{amsmath}
\usepackage{lastpage}
	\usepackage{pdflscape}

	\usepackage{enumerate}
	\usepackage{makeidx}
	\usepackage{amsfonts}
	\usepackage[ansinew]{inputenc}
	\usepackage[usenames,dvipsnames]{pstricks}
	\usepackage{subfigure}
	\usepackage{epsfig}
	\usepackage{pst-grad} 
    \usepackage[T1]{fontenc} 
	\usepackage[colorlinks,hyperindex]{hyperref}
	\hypersetup
	{
		colorlinks,%
		citecolor=black,%
		linkcolor=black,%
		urlcolor=black,%
	}
    \usepackage{setspace}
	
\newcounter{concno}
\newcommand{\conc}[1]{\refstepcounter{concno} \Alph{concno}\label{#1}}

\newcommand{\sqb}[1]{\left[ #1 \right]}

\newcommand{\be}{\begin{equation}}
\newcommand{\ee}{\end{equation}}
\newcommand{\bea}{\begin{eqnarray}}
\newcommand{\eea}{\end{eqnarray}}

\newcommand{\ba}{\begin{array}}
\newcommand{\ea}{\end{array}}

\newcommand{\refer}[1]{(\ref{#1})}
\newcommand{\dd}{\mbox{d}}

\newcommand{\Lag}{\mathcal{L}}
\providecommand{\fp}[2]{\frac{\partial {#1}}{\partial {#2}}} 

\begin{document}


\fancyhead[c]{\small To be added by the Editor
 } \fancyfoot[C]{\small To be added by the editor-\thepage}
\footnotetext[0]{To be added by the editor}

\title{Existence and stability of circular orbits in static and axisymmetric spacetimes\thanks{The work of J. Jia and X. Pang are supported by the NNSF China 11504276 \& 11547310 and MOST China 2014GB109004. The work of N. Yang is supported by the NNSF China 31401649 \& 31571797.}}

\author{Junji Jia$^{1,2;1)}$\email{junjijia@whu.edu.cn} \and Xiankai Pang$^{1,2}$ \and Nan Yang$^{3;2)}$\email{nanyang@hbut.edu.cn}}

\maketitle

\address{%
$^1$ MOE Key Laboratory of Artificial Micro- and Nano-structures, Wuhan University, 430072, China\\
$^2$ Center for Theoretical Physics, School of Physics and Technology, Wuhan University, Wuhan, 430072, China\\
$^3$ Glyn O. Phillips Hydrocolloid Research Centre, Hubei University of Technology, Wuhan 430068, China}

\begin{abstract}
The existence and stability of timelike and null circular orbits (COs) in the equatorial plane of general static and axisymmetric (SAS) spacetime are investigated in this work. Using the fixed point approach, we first obtained a necessary and sufficient condition for the non-existence of timelike COs. It is then proven that there will always exist timelike COs at large $\rho$ in an asymptotically flat SAS spacetime with a positive ADM mass and moreover, these timelike COs are stable. Some other sufficient conditions on the stability of timelike COs are also solved. We then found the necessary and sufficient condition on the existence of null COs. It is generally shown that the existence of timelike COs in SAS spacetime does not imply the existence of null COs, and vice-versa, regardless whether the spacetime is asymptotically flat or the ADM mass is positive or not. These results are then used to show the existence of timelike COs and their stability in an SAS Einstein-Yang-Mills-Dilaton spacetimes whose metric is not completely known. We also used the theorems to deduce the existence of timelike and null COs in some known SAS spacetimes.
\end{abstract}



\footnotetext[0]{\hspace*{-3mm}\raisebox{0.3ex}{$\scriptstyle\copyright$|}To be added by the editor}%

\begin{multicols}{2}

\section{Introduction}

Axisymmetric spacetimes are of special interests in observational astrophysics and theoretical study of spacetime and gravitational theories. This symmetry is crucial to some important observable phenomena such as torus or jet-like features \cite{DelZanna:2004aq}, fast rotating objects, supernova ejecta \cite{Wang:2002rx} and binary systems that generates gravitational waves. On the other hand, many theoretical work such as Einstein-Yang-Mills (EYM) \cite{Radu:2001ij}, EYM-Higgs \cite{Hartmann:2001ic}, EYM-Dilaton \cite{Kleihaus:1996vi,Kleihaus:1997ws,Kleihaus:1997mn} solutions, f(R) gravity \cite{Capozziello:2009jg}, traversable wormholes
\cite{Kuhfittig:2003wr}, cosmic strings \cite{Reddy:2006np} and domain walls \cite{Reddy:2006dz}, and metric affine gravity \cite{Vlachynsky:1996zh} can be carried out in this symmetry due to its simplicity and yet non-triviality. In particular, the recent direct observation of gravitational waves \cite{Abbott:2016blz,Abbott:2016nmj} are believed to originate from binary system whose spacetime in the far region during merging and spacetime in the entire space after complete merging, are indeed axisymmetric.

In the spacetimes possessing axisymmetry, it is often desirable to know the existence and stability of circular orbits (COs) in the equatorial plane. These orbits for example can be used to characterize the properties of the central object
\cite{Hackmann:2011wp} or in approximate treatment of more complicated equatorial motion \cite{SanabriaGomez:2010vb} and numerical modeling of stellar systems \cite{Thomas:2004rz}. They are also of huge importance to the study of flow of matter in the accretion disk around rotating black holes (BHs) or binary systems \cite{Shibata:1998xw,Donati:2005tw,Abramowicz:2010nk}. For these reasons, these COs, including the ones that are more special such as marginal stability COs (MSCOs), have been studied by many authors over the years for various axisymmetric spacetimes, including both static and stationary ones (see Ref. \cite{Hackmann:2011wp} and papers therein).

In this work, we study the existence and stabilities of COs in the equatorial plane of general static and axisymmetric (SAS) spacetimes. This work is a natural extension of the authors' recent results on the existence and stability of COs in general spherically symmetric spacetimes \cite{Jia:2017nen}. Previously, there have been work on the solution of COs and their stabilities in {\it particular} SAS spacetimes. Letelier \cite{Letelier:2003ea} studied the stability of BH + disk/ring/multipolar systems using the Rayleigh criteria. Gonz\'{a}lez and L\'{o}pez-Suspes considered the stability of COs in Weyl (vacuum) spacetimes \cite{Gonzalez:2011fb}. Dolan and  Shipley studied the stable photon orbits in a few spacetimes \cite{Dolan:2016bxj}. In addition, Beheshti and Gasper\'{\i}n studied the MSCO in particular stationary and static axisymmetric spacetimes \cite{Beheshti:2015bak}. We emphasis that unlike works by these authors, in this work the metric function of the SAS spacetimes are kept general, i.e., the results that we obtained are not restricted to a particular forms of metric functions. Rather, our methodology deals with general SAS spacetimes and our results are even applicable to metrics whose explicit expression is not yet known.

We organize the paper in a few sections. In section \ref{metmin} we setup the SAS metric, and derive the geodesic equations in the equatorial plane and the CO conditions in terms of the metric functions and first integral constants. In section \ref{timelikecase} we analyze the existence of non-trivial timelike COs and give the results in the form of a theorem. We then show in section \ref{sectimeasy} that asymptotically flat SAS spacetimes with a positive ADM mass will always allow timelike COs. The stabilities of these timelike COs are studied and some sufficient conditions for the COs to be stable (or unstable) are obtained. The existence and stability issue for null CO are studied in section \ref{nullcase}. It is found that there can exist metrics allowing timelike CO but no null COs, and metrics that allowing null CO but no timelike COs.
Finally, in section \ref{discuss} we show  a very powerful application of our results to assert the existence and stability of COs for SAS EYMD spacetime whose metric functions' exact expressions are not known. We also use our results to study the existence of COs in a few other known SAS spacetimes and discuss possible extensions of the current work.

\section{Metric, geodesic equations and CO conditions\label{metmin}}

In this section, we setup the metric and derive the geodesic equations and conditions for the COs in the equatorial plane of a stationary and axisymmetric spacetime, although what we need in later sections are only these quantities in a {\it static} and axisymmetric spacetime. Our notations in this section follow that of Ref. \cite{Beheshti:2015bak}.

For stationary and axisymmetric spacetimes, the most general form of the metric can be given in local coordinates by the Weyl-Lewis-Papapetrou line element \cite{Beheshti:2015bak}
\be
   \dd s^2=e^{2U}(\dd t-\omega \dd \phi)^2-e^{-2U}[e^{2\gamma}(\dd \rho^2+\dd z^2)+\rho^2\dd \phi^2]\label{metric0}
\ee
where $(t,~\rho,~z,~\phi)$ are the coordinates and $U,~\omega$ and $\gamma$ are the metric functions depending on $\rho$ and $z$ only. Equivalently, and for easier reference to the metric functions, we can rewrite it as
\bea
\dd s^2&=& A(\rho,z)\dd t^2-B(\rho,z)\dd t\dd \rho \nonumber \\
   &&-C(\rho,z)\dd \phi^2-D(\rho,z)(\dd \rho^2+\dd z^2) \label{metric}
\eea
where we have set $ A(\rho,z)=e^{2U}$, $B(\rho,z)=2 \omega e^{2U}$, $C(\rho,z)=\rho^2e^{-2U}-\omega^2e^{2U}$ and $D(\rho,z)=e^{2(\gamma-U)}$. Note there is a relation between these functions
\be C=\left(\rho^2-B^2/4\right)/A. \label{cinab}\ee
We further assume that the spacetime possesses a local reflection symmetry at some fixed $z$ value and we shift the $z$ coordinate so that it becomes the $z=0$ plane, which is also called the equatorial plane. This way any orbit of a particle with zero initial off-plane momentum will remain in the plane and the motion becomes effectively 2+1 dimensional whose metric is described after changing from $A(\rho,z)$ etc. to $A(\rho)$ etc. by
\be
   ds^2=A(\rho)\dd t^2-B(\rho)\dd t\dd \rho -C(\rho)\dd \phi^2-D(\rho)\dd \rho^2. \label{metric2}
\ee

To obtain the geodesic equations, we can start from the following Lagrangian of a free particle
\bea
   \Lag&=&\frac{1}{2}g_{ik}\frac{dx^i}{d\tau}\frac{dx^k}{d\tau}=\frac{1}{2}\epsilon \nonumber \\
				   &=&\frac{1}{2} \left[\frac{B\dot{\phi}^2}{4
 A}-\frac{\rho ^2 \dot{\phi}^2}{A}+ A\dot{t}^2 -B \dot{t} \dot{\phi}- D\dot{\rho}^2\right] \label{lagdef}
\eea
where $\dot{}$ denotes the derivative with respect to the proper time (or affine parameter) $\tau$ and we have substituted \refer{cinab}. Here $\epsilon=1,~0$ respectively for timelike and null geodesics. Because $t$ and $\phi$ are cyclic coordinates in Lagrangian, we can obtain two first integrals
 \bea
   E&=&\fp{\Lag}{\dot{t}}=\frac{1}{2} \left(2
 A(\rho ) \dot{t}-B(\rho ) \dot{\phi}\right), \label{Edef}\\
 L&=&\fp{\Lag}{\dot{\phi}}=\frac{1}{2}
   \left[\frac{1}{2A(\rho)} \left(B(\rho )^2-4\rho ^2\right) \dot{\phi}-B(\rho )
   \dot{t}\right], \label{Ldef}
\eea
where $E$ and $L$ are real constants identified as the specific energy and angular momentum of the particle at infinite $\rho$. These equations can be inverted to express $\dot{t}$ and $\dot{\phi}$ in terms of $E$ and $L$ as
 \bea
   \dot{t}&=&\frac{E}{A(\rho)}\left(1-\frac{ B(\rho )^2}{4 \rho ^2 }\right)-\frac{L B(\rho )}{2 \rho ^2} , \label{tdotsol}\\
   \dot{\phi}&=&-\frac{L A(\rho )}{\rho ^2}-\frac{E B(\rho )}{2 \rho ^2} .\label{phidotsol}
\eea
These are equivalent to the geodesic equations of $t$ and $\phi$ but with $\tau$ integrated once.
Substituting Eqs. \eqref{tdotsol} and \eqref{phidotsol} into Lagrangian \refer{lagdef} yields the last geodesic equation, for $\rho(\tau)$, as
\bea
   \dot{\rho}^2&=&\frac{\left[-4A\left(E L B +\epsilon  \rho ^2 \right)  -4 L^2 A^2 +E^2 \left(4 \rho ^2-B^2\right)\right] }{4 \rho ^2 A  D} \label{Vdef1}\\
		&\equiv& \frac{\Phi(\rho)}{4\rho^2A(\rho)D(\rho)} \equiv  V(\rho) \label{Vdef}
\eea
where $\Phi(\rho)$ is defined as the numerator and $V(\rho)$ as the entire right hand side of Eq. \eqref{Vdef1} and plays the role of an effective potential.

To establish the condition for a CO, we can first compute the conjugate momentum $p_\rho$ of $\rho$ as
\bea
  p_\rho&=&\fp{\Lag}{\dot{\rho}}=-D(\rho)\dot{\rho} \label{prhodef}
\eea
Taking derivative with respect to $\tau$, we obtain
 \bea
   \dot{p_\rho}&=&\frac{d}{d\tau}(-D(\rho)\dot{\rho}) =-\dot{\rho}\frac{d}{\dd \rho}(D(\rho)\dot{\rho}) \nonumber \\											   &=&-\frac{1}{2D(\rho)}\frac{d}{\dd \rho}(D^2(\rho)\dot{\rho}^2) \nonumber \\
&=&-D(\rho)\left(D^\prime(\rho)V(\rho)-\frac{V^\prime(\rho)}{2}\right) \label{eqprhodot}
\eea
where in the last step we used Eq. \refer{Vdef} and $^\prime$ here and henceforth denotes the derivative with respect to $\rho$.
Mathematically, Eqs. \eqref{prhodef} and \eqref{eqprhodot} can be thought as an autonomous system in 2-dimensional phase space spanned by $\rho$ and $p_\rho$. Then the fixed point (FP) of this system, denoted by $(\rho_*,~p_{\rho*})$, satisfies
\bea
\left\{
\begin{aligned}
  \dot{\rho}|_{(\rho_*,p_{\rho*})}&=-\frac{p_{\rho*}}{D(\rho_*)}=0, \\
   \dot{p_\rho}|_{(\rho_*,p_{\rho*})}&=-D(\rho_*)\left[D^\prime(\rho_*)V(\rho_*)-\frac{V^\prime(\rho_*)}{2}\right]=0
\end{aligned}\label{autosys}
\right.
\eea
at some instantaneous time and all times after. Noting \eqref{Vdef}, this is equivalent to require that at $(\rho_*,~p_{\rho*})$
 \bea
   \Phi(\rho_*)=0\quad \text{~and} \quad \Phi^\prime(\rho_*)=0 .\label{fpcon}
\eea
Clearly any solution of Eq. \refer{autosys} or equivalently \refer{fpcon} defines a CO of the spacetime with orbit radius given by $\rho_*$. Conversely, it is also easy to show that any CO in this spacetime will satisfy these equation systems. Therefore COs of the metric \refer{metric2} and FP defined by the system \refer{autosys} or \refer{fpcon} are equivalent.

For this work, we then restrict ourselves to the case of static metric, mainly due to its simplicity and the solvability of the desired equations. This allows us to set $B(\rho)=0$ in all the equations above, including metric \refer{metric2} and Eqs. \refer{autosys} and \refer{fpcon}. These two systems then will be our starting point for the analysis of the existence and stability of COs in the spacetime described by metric
\be
   ds^2=A(\rho)\dd t^2 -\rho^2/A(\rho)\dd \phi^2-D(\rho)\dd \rho^2. \label{metric3}
\ee

\section{Existence of timelike CO\texorpdfstring{\MakeLowercase{s}}{} \label{timelikecase}}

We consider the existence of COs for timelike geodesics in this section, while the null case will be considered in section \ref{nullcase}.

First we write out Eq. \eqref{fpcon} explicitly in terms of the parameters $E,~L$ and the metric function $A(\rho)$, as
\be
\left(
\ba{cc}
\rho_*^2              & -A^2\\
\rho_*^3A^\prime&[\rho_* A^\prime-2A]A^2
\ea
\right)
\left(
\ba{c}
E^2\\
L^2
\ea
\right)=
\left(
\ba{c}
\rho_*^2 A\\
0
\ea
\right).\label{eqmat}\ee
For a given $A(\rho)$, if in the space spanned by energy $E$ and angular momentum $L$ there exists a non-empty set $S$ for whose element $(E,~L)$ the solution to Eqs. \refer{eqmat} does exist, then we say for that $A(\rho)$ the COs can exist (for that $(E,~L)$ at $\rho=\rho_*$). If the set $S$ is empty then we say that there exist no COs for the spacetime described by that $A(\rho)$. Apparently, if COs exist for some $(E,~L)$ then the two equations in Eq. \refer{eqmat} will both have solutions simultaneously. If either
of the equations is not satisfied by any $\rho$, then the COs do not
exist for that $(E,~L)$. Therefore the form of $A(\rho)$ is crucial in determining the existence of COs.
Before proceeding further however, a remark is in order regarding $A(\rho)$. Even though it is possible for $A(\rho)$ to be negative in some region of the spacetime, in this paper we assume that $A(\rho)$ is positive in the range of $\rho$ in which we seek the FP (or CO), because otherwise according to the metric \refer{metric3} the $\rho$ coordinate should have to be interpreted as time, not radius and a FP in time is not of our interest. We also assume that in the range of $\rho$ the spacetime is not singular so that it is always possible to do coordinate transformation to make $A(\rho)$ differentiable if it was not in the first place. Therefore throughout this paper, we will assume that $A(\rho)$ is already made differentiable.

Eq. \refer{eqmat} is a linear system for $E^2$ and $L^2$. It will have no solution to $E^2$ and $L^2$ if and only if the rank of the augmented matrix is larger than that of the coefficient matrix, which we solved to find
\be
A(\rho_*)-\rho_* A^\prime(\rho_*)=0. \label{ranknosol}\ee
Otherwise, there always exist the following solution
\bea
   E^2&=&\frac{ A(\rho_* ) \left[2 A(\rho_* )-\rho_* A^\prime(\rho_* )\right]}{2 \left(A(\rho_*)-\rho_*  A^\prime(\rho_*)\right)} ,\label{Esolsta} \\
   L^2&=&\frac{\rho_* ^3  A^\prime(\rho_* )}{2 A(\rho_*)\left(A(\rho_*)-\rho_*  A^\prime(\rho_*)\right)} .\label{Lsolsta}
\eea
Given that both $E$ and $L$ are real, it is clear that these two equations might not have solutions if the right hand sides are always negative for all $\rho$. However, it is also clear that if $A(0)>0$, there will always exist a solution at $\rho_*=0$ for $E=\sqrt{A(0)}$ and $L=0$. This solution, though technically is also a CO, is of less interests in most physically important metrics. Usually in the center $\rho=0$, either the spacetime is singular, or the presence of matter will prevent the test particle from doing timelike geodesic motions if this particle is not interacting weakly with matter, or it is just a trivial CO such as in the case of Minkovski spacetime. Therefore in the following sections when the timelike COs are discussed, CO at this point is excluded and we concentrate on non-trivial COs.

We can now study Eq. \refer{Esolsta} and \refer{Lsolsta}
separately for the (non-)existence of their solutions.
Firstly for Eq. \refer{Esolsta}, keeping in mind that $E$ is real and noticing $(2A-\rho A^\prime)-(A-\rho A^\prime)=A>0$, this equation will not have any solution in a region of $\rho$ if and only if in this  region
\bea && A-\rho A^\prime<0\\
\mbox{and }&&2A-\rho A^\prime>0 .\eea
That is, in this region of $\rho$ we can equate the left hand sides of the inequalities to some arbitrary but positive functions $\delta(\rho)$ and $\zeta(\rho)$:
\bea && A(\rho)-\rho A^\prime(\rho)=-\delta(\rho),\label{eqecd1}\\
\mbox{and }&& 2A(\rho)-\rho A^\prime(\rho)=\zeta(\rho).\label{eqecd2}\eea
Under condition \refer{eqecd1} alone we see that $A^\prime>0$ and consequently Eq. \refer{Lsolsta} will not have any solution. Therefore for the purpose of violating the entire system of \refer{Esolsta} and \refer{Lsolsta}, condition \refer{eqecd1} is enough and we do not have to solve \refer{eqecd2}. Noticing Eq. \refer{ranknosol}, the $\delta(\rho)$ in Eq. \refer{eqecd1} can be relaxed to be semi-positive rather than strictly positive. This equation can be readily solved for an arbitrary $\delta(\rho)$. However for the sake of a shorter expression for the solution, we further change $\delta(\rho)$ to $\rho^2\dd \kappa(\rho)/\dd \rho$ without losing any generality as long as $\kappa^\prime(\rho)\geq0$. This way the Eq. \refer{eqecd1} can be solved to find the solution
\be
  A(\rho)=\rho \kappa(\rho) \label{Asol}
\ee
where an integral constant has been absorbed into $\kappa(\rho)$, which is a positive and monotonically increasing but otherwise arbitrary function.

Secondly, for Eq. \refer{Lsolsta}, beside the case \refer{eqecd1}, this equation can also be violated if its numerator is negative
\be A^\prime(\rho)<0, \label{eqlcd}\ee which automatically makes the denominator positive. It is also easy to see that in all other cases Eq. \refer{Lsolsta} will have solutions for some $(E,~L,~\rho)$.

Combining the above two cases, we see that the equation system \refer{Esolsta} and \refer{Lsolsta} will not permit a CO solution in a region of $\rho$ if and only if at least one of Eq. \refer{Asol} and \refer{eqlcd} is satisfied in this region. If no CO exists at all for all $\rho>0$, then in the entire range of $\rho$, either one of \refer{Asol} or \refer{eqlcd} is satisfied, or these two are satisfied piecewisely.   Now we further show that the piecewise satisfaction scenario is impossible and then establish one of the main conclusions in this work.

Suppose that there exist no CO because Eq. \refer{Asol} or equivalently Eq. \refer{eqecd1},  and \refer{eqlcd} are piecewisely satisfied: Eq. \refer{eqecd1} is satisfied in region $\rho \in (a,~b)$ and Eq. \refer{eqlcd} is satisfied in the region $\rho\in(b,~c)$. Then clearly $A^\prime>0$ in $(a,~b)$ and in $A^\prime<0$ in $(b,~c)$. Because $A^\prime$ is continuous, we must have $A^\prime(\rho=b)=0$ and for any $\rho$ in the small neighborhood $\rho\in(b-\chi,~b)$,  $A^\prime(\rho)>0$. On the other hand for $\rho=b-\varepsilon$ which is in this neighborhood and infinitesimally close to $b$, we can compute $[A-\rho A^\prime]_{\rho=b-\varepsilon}$ using its Taylor expansion
\be
  [A-\rho A^\prime]_{\rho=b-\varepsilon}=A(b)+\varepsilon bA^{\prime\prime}(b)+{\cal O}(\varepsilon^2).
\ee
Clearly we have $[A-\rho A^\prime]_{\rho=b-\varepsilon}=A(b)>0$ to the leading order, which conflicts with the assumption that Eq. \refer{eqecd1} is satisfied here. For the case that Eqs. \refer{eqecd1} and \refer{eqlcd} are satisfied respectively in the regions $(b,~c)$ and $(a,~b)$, then a similar conflict can be proven. These conclude that Eqs. \refer{Asol} and \refer{eqlcd} cannot be satisfied piecewisely. We can now state our conclusion:\\
(Theorem \conc{nopiece}) The equatorial timelike COs for the metric \refer{metric} do not exist if and only if either only Eq. \refer{Asol} is satisfied in the entire range of $\rho$, or only Eq. \refer{eqlcd} is satisfied in the entire range of $\rho$, but not because Eq. \refer{Asol} and \refer{eqlcd} are piecewisely satisfied.\\
Here and after, we will refer to this as the timelike CO non-existence theorem (TCONET).

In the following we give two simple examples satisfying respectively Eq. \refer{Asol} and \refer{eqlcd} in the entire range of $\rho$ and therefore do not permit any CO, and one more example which combines the first two and satisfies Eq. \refer{Asol} in some range of $\rho$, Eq. \refer{eqlcd} in some other range and then neither of them in the rest, where CO lies. The first example is
given by $\kappa(\rho)=\rho$ so that $A(\rho)=\rho\kappa(\rho)=\rho^2$, the resulting Eqs. \eqref{Esolsta} and \eqref{Lsolsta} become
\bea
E^2=0,\quad L^2=-1
\eea
which permit no solution and therefore no COs. The second example is given by $A(\rho)=1/\rho$ and the Eqs. \eqref{Esolsta} and \eqref{Lsolsta} become
\bea
  E^2=\frac{3}{4\rho},\quad L^2=-\frac{\rho^3}{4}
\eea
where the equation for $L^2$ eliminates the existence of COs too.
The third example is a simple combination of the above two:
\be
A(\rho)=\frac{\rho}{1+\rho^3}\cdot\rho. \label{egA3}
\ee
At very small and large $\rho$, this resembles the first and second examples respectively and therefore we expect that the COs will not exist there. However, in the middle range, as we can expect using the TCONET theorem, the COs do exists. Substituting Eq. \eqref{egA3} into \eqref{Esolsta} and \eqref{Lsolsta}, we obtain
\bea
E^2=\frac{3 \rho ^5}{2 \left(2 \rho ^6+\rho ^3-1\right)},~L^2=\frac{-\rho ^6+\rho ^3+2}{4 \rho ^3-2}.
\eea
It is easy to verify that these two equations will have solution when $\rho\in(2^{-1/3},2^{1/3})$ for some $(E,~ L)$.

\section{EXISTENCE OF CO\texorpdfstring{\MakeLowercase{s}}{} in ASYMPTOTICALLY FLAT SAS SPACETIMES \label{sectimeasy}}

In the end of last section, we see that there indeed exist metrics forbidding the existence of any timelike CO. However, these metrics are not asymptotically flat. Usually asymptotically flat spacetimes are more interesting due to their physical relevance. Therefore in this section we analyze how the asymptotic flatness condition will affect the existences of timelike COs in the SAS spacetimes.

For metric \refer{metric2}, asymptotic flatness requires that as $\rho\to\infty$ \cite{Schmidt:2000rta,Semerak:2002uc},
\be
A(\rho)=a+\frac{2c}{\rho^\beta}+{\cal O}(\rho^{-\beta-1}),\label{asyreq}
\ee
where $a>0,~\beta\geq1$ and $c$ are constants.
We then see that any $A(\rho)$ satisfying \refer{Asol} will necessarily diverge at large $\rho$ and therefore not be asymptotically flat. While for condition \refer{eqlcd}, metric satisfying it can still be compatible with the asymptotic flatness condition \refer{asyreq} and therefore has no timelike COs. An example would be the function \refer{asyreq} truncated to the second term and with a positive $c$. Clearly this metric satisfies
\be A^\prime=-\frac{2 \beta c}{r^{\beta+1}}<0, \ee
i.e. Eq. \refer{eqlcd} and therefore has no timelike CO. This example shows that enforcing the asymptotic flatness alone will not make sure the existence of COs in SAS spacetimes.

There is however, one thing usual about this example. The constant $c$ in \refer{asyreq} although is mathematically merely a constant in the asymptotic expansion of the metric function $A(\rho)$, however physically it is identified with the ADM mass for the asymptotically flat spacetime when $\beta=1$. It is known from the Positive Mass Theorem that for spacetime satisfying dominant energy condition, this ADM mass shall always be positive \cite{Mars:2014wbd}.
Therefore adding the positive ADM mass condition will make sure that the metric function $A(\rho)$ violate Eq. \refer{eqlcd} and guarantee the existence of COs in SAS spacetimes. We restate this conclusion as:\\
(Theorem \conc{asyflatexistence}) There will always exist equatorial timelike COs at least at large $\rho$ for an SAS spacetime described by metric \refer{metric} and which is also asymptotically flat with a positive ADM mass.

\section{Stability of CO\texorpdfstring{\MakeLowercase{s}}{} and existence of marginally stable CO\texorpdfstring{\MakeLowercase{s}}{} \label{sectimestab}}

With the existence of COs proven for asymptotically flat SAS spacetimes with positive ADM mass, and existence condition given by Theorem \ref{nopiece} for general SAS spacetimes, the next important question is the stability of these COs. In this section we will assume that the spacetime considered will allow timelike CO and denote its radius as $\rho=\rho_*$. We then use the Lyapunov exponent method \cite{Pradhan:2016} to address the question about what form of $A(\rho)$ will make the CO stable, unstable or have a marginal stability. Here we use the wording ``marginal stability'' instead of ``marginally stable'' \cite{Beheshti:2015bak} or ``marginal stable'' \cite{Ono:2016lql} because such orbit, defined as $\dd V(\rho_*)/\dd \rho =0$, is not necessarily stable even in the perturbative sense. This definition only means that the stability at this point is critical, i.e., it can change, disappear or split in to many equilibria \cite{mscodef}.

For the metric \eqref{metric3}, the Lyapunov exponents for the system \refer{autosys} (or \refer{fpcon}) is equivalent to the eigenvalues of the linear perturbation matrix of the system, which we found to be
\be
  \lambda_\pm=\pm\sqrt{\frac{V^{\prime\prime}(\rho_*)}{2}} ,\ee
where the effective potential $V(\rho)$ is given in Eq. \refer{Vdef} with $B=0$. Substituting $V$ and using the relation \refer{Esolsta} and \refer{Lsolsta} at the CO yields
\bea
\lambda_\pm&=&\pm\sqrt{\frac{\rho_* A^2 A^{\prime\prime}+\rho_*^2 A^{\prime3}-4\rho_* A A^{\prime2}+3A^2 A^\prime }{2\rho_*  A^2 D (\rho_*  A^\prime-A)}}\nonumber \\
&\equiv&\pm\sqrt{\frac{h(\rho_*)}{2\rho_*  A^2 (\rho_*  A^\prime-A)D}}
\eea
where all of $A,~A^\prime,~A^{\prime\prime}$ and $D$ are evaluated at $\rho_*$, and we defined a function $h$ as the numerator
\be
  h(\rho)= \rho A^2 A^{\prime\prime}+\rho^2 A^{\prime3}-4\rho A A^{\prime2}+3A^2 A^\prime  .\label{def_h}
\ee
It is known that when $\lambda_\pm$ are imaginary, real but nonzero, or zero, the CO will be respectively stable, unstable, or have marginal stability. Since at the CO radius, the Eq. \eqref{eqecd1} must be broken and therefore $\rho_*A^\prime(\rho_*)-A(\rho_*)<0$, it is clearly then the sign of $\lambda_\pm$ is completely determined by $h$. I.e., the CO will (1) be stable if and only if $h(\rho_*)>0$, (2) be unstable if and only if $h(\rho_*)<0$, and (3) have marginal stability if and only if $h(\rho_*)=0$.
Unfortunately, the sign of $h(\rho_*)$ cannot be completely determined without knowledge of $A(\rho)$, due to two facts.  First, the $A^{\prime\prime}(\rho_*)$ term, which does not appear in any form of the CO existence conditions, is not constrained and cannot be eliminated from \refer{def_h}. Secondly, even if $A^{\prime\prime}(\rho_*)$ were expressible through other quantities, $h(\rho_*)$ is only an algebraic equation at $\rho_*$. Without knowing the value of the CO radius $\rho_*$ it is still impossible to determine the sign of $h(\rho_*)$.

However, if we treat $h(\rho)$ as a differential equation of $\rho$ and require its sign to be fixed at not only $\rho_*$ but all $\rho$, then we are still able to get some sufficient conditions for determining the stability of COs of relevant metrics. These conditions are:\\
(a). If
\be \rho A^2 A^{\prime\prime}+\rho^2 A^{\prime3}-4\rho A A^{\prime2}+3A^2 A^\prime =0\label{v2ca}\ee
for all $\rho$, then the CO will always be an MSCO regardless of the CO radius value $\rho_*$. \\
(b). If
\be \rho A^2 A^{\prime\prime}+\rho^2 A^{\prime3}-4\rho A A^{\prime2}+3A^2 A^\prime =\sigma(\rho)>0~ (\mbox{or}~<0) \label{v2cb}\ee
where $\sigma(\rho)$ is positive (or negative) but otherwise arbitrary for all $\rho$, then the CO will always be an stable (or unstable) CO, regardless of the CO radius value $\rho_*$.

Condition (a) can be immediately solved to get two possible $A(\rho)$
\be A_\pm(\rho)=\rho\left(c_1\rho\pm \sqrt{c_1^2\rho^2-2c_2}\right), \ee
where for the $+$ sign, we should have $c_1>0$ or ($c_1<0, ~c_2<0$) and for the $-$ sign, $c_1>0 and~c_2>0$ in order for $A(\rho)$ to be positive. The relations \refer{Esolsta} and \refer{Lsolsta} at the CO radius then further eliminate the scenario of $A_+$ with $c_1>0$ and $A_-$ since they do not allow any CO. Combining these, then the condition (a) is equivalent to the statement that:\\
(Theorem \conc{mscothm}) For an SAS spacetime with metric function $A(\rho)=\rho\left(c_1\rho+ \sqrt{c_1^2\rho^2-2c_2}\right)$ where $c_1$ and $c_2$ are negative but otherwise arbitrary constants, there will exist timelike COs and these COs are always MSCOs.

Condition (b) allows better generality than (a) due to the arbitrariness of $\sigma(\rho)$ but has proven to be difficult to solve. This difficulty on one hand is due to the nonlinearity of the equation, and on the other due to the non-homogeneity of $\sigma(\rho)$. Therefore we tried to replace $\sigma(\rho)$ by a homogenous term, such as $\chi(\rho) \rho^2A^{\prime3}$, $\chi(\rho) \rho AA^{\prime2}$ or $\chi(\rho) A^2A^\prime $, where $\chi(\rho)$ is also an arbitrary function that has the same sign as $\sigma(\rho)$. This way since the other factors in these terms are all positive, if the resulting equations were solvable, we can still obtain some sufficient conditions with enough generality on the stability of the COs. Again it is very unfortunate that these equations do not allow explicit solutions when $\chi(\rho)$ is an arbitrary function. At last, we further set $\chi(\rho)$ to be constants $\chi$, and the resultant equations became simple enough to solve and each solved $A(\rho)$ is a sufficient conditions for the CO to be stable or unstable, depending on the sign of $\chi$. The detailed forms of these $A(\rho)$'s are quite implicit and not directly useful, and therefore we only list them in Appendix \ref{appda}.

Similar to the situation in section \ref{sectimeasy}, one can also tackle the stability problem of timelike COs at large $\rho$ with the help of asymptotic flatness and positive ADM mass. From Theorem \ref{asyflatexistence}, we see the existence of timelike COs at large $\rho$ for metrics having asymptotics \refer{asyreq} with $c<0$ and $\beta=1$. For these COs, substituting $A(\rho)$ into Eq. \refer{def_h}, we find
\be h(\rho)=\frac{a^2c\beta(\beta-2)}{\rho^{\beta+1}}+{\cal O}(\rho^{-(\beta+2)}), \ee
which is positive to the leading order. This means we can have the following result:\\
(Theorem \conc{flatplusunstab}) The equatorial timelike COs at large $\rho$ in the asymptotically flat SAS spacetime with positive ADM mass are always stable.

To conclude this section, let us mention that an especially simple sufficient condition for the COs to be stable can be obtained from \refer{def_h} just by observation. Noticing that $h(\rho)$ can be recast into
\be
h(\rho_*)=A^\prime \left(A-\rho_*  A^\prime\right) \left(3 A-\rho_* A^\prime\right)+\rho_* A^2A^{\prime\prime}\label{hsecondform}\ee
and that the three factors in the first term of the right hand side are all positive, we see that as long as the last term is semi-positive definite, then the COs will be unstable. That is, the COs will be stable if $A^{\prime\prime}(\rho_*)\geq0$.

\section{Null CO\texorpdfstring{\MakeLowercase{s}}{} and their stabilities \label{nullcase}}

For null geodesics, the existence conditions \eqref{fpcon} for the null COs becomes
\bea
  \frac{E}{L}&=&\pm\frac{A(\rho_*)}{\rho_*}, \label{nc1eq1}\\
  A(\rho_*)&=&\rho_*A^\prime(\rho_*). \label{nc1eq2}
\eea
It is clear that Eq. \refer{nc1eq1} will always be satisfied by some $\rho, ~E$ and $L$. Eq. \refer{nc1eq2} will have no solution and therefore no COs if and only if for all $\rho>0$, we have either
\be
A(\rho)-\rho A^\prime(\rho)=\eta_1(\rho)>0, \label{fpcon_A_null1}
\ee
or
\be
A(\rho)-\rho A^\prime(\rho)=-\eta_2(\rho)<0 \label{fpcon_A_null2} ,\ee
where $\eta_1(\rho)$ and $\eta_2(\rho)$ are two positive arbitrary functions.
Noticing that $A(\rho)$ and $A^\prime(\rho)$ are both continuous, it is seen that the Eqs. \refer{fpcon_A_null1} and \refer{fpcon_A_null2} cannot be satisfied piecewisely in the range of $\rho$ without letting $A(\rho)-\rho A^\prime(\rho)$ pass zero, i.e., having a CO.

Eqs. \refer{fpcon_A_null1} and \refer{fpcon_A_null2} have the same form as Eq. \refer{eqecd1} and the solutions are respectively
\be
A(\rho)=\rho\psi_1(\rho) \label{nstaticsol1}\ee
where $\psi_1(\rho)$ is a positive and monotonically increasing but otherwise arbitrary function,  and
\be
A(\rho)=\rho\psi_2(\rho) \label{nstaticsol2} \ee
where $\psi_2(\rho)$ is an positive and monotonically decreasing but otherwise arbitrary function.
In other words, we have:\\
(Theorem \conc{nstaticthm}) The equatorial null COs for the metric \refer{metric} do not exist if and only if its metric function $A(\rho)$ takes the form of either Eq. \refer{nstaticsol1} or Eq. \refer{nstaticsol2} in the entire range of $\rho$.

Comparing with the TCONET,  any metric function $A(\rho)$ satisfying Eqs. \refer{eqlcd} will satisfy Eq. \refer{fpcon_A_null1} or equivalently Eq. \refer{nstaticsol1}. However metric satisfying \refer{eqecd1} will not satisfy either of Eqs. \refer{fpcon_A_null1} and \refer{fpcon_A_null2} when the equality $\delta(\rho)=0$ or $\kappa^\prime(\rho)=0$ is taken. This means that there are SAS spacetimes that does not have timelike COs but still allow null COs.
An example can be constructed as
\bea
&& \kappa(\rho)=a^3+(\rho-a)^3~~~(a>0),\nonumber\\
&& A(\rho)=\rho\kappa(\rho)=\rho\left[a^3+(\rho-a)^3\right], \eea
which leads to $\kappa^\prime(r)=3(\rho-a)^2\geq0$ and therefore $A(\rho)$ respects Eq. \refer{Asol} and has no timelike CO. It is also clear that at $\rho=a$, $A-\rho A^\prime=0$ and therefore there exist a null CO. On the other hand, one can also give examples of metrics that allow timelike COs but not null CO, such as
\be \psi_2(\rho)=\frac{1}{\rho+2M},~A(\rho)=\rho\psi_2(\rho)=1-\frac{2M}{\rho+2M} \label{theeg}\ee
where $M>0$ is the ADM mass. The $\psi_2(\rho)$ here is positive and has a negative derivative and therefore will have no null CO. It is also easy to verify that $A(\rho)$ violates both conditions \refer{Asol} and \refer{eqlcd} and therefore allows timelike COs.

We can impose the asymptotic flatness condition to the metric. Similar to the timelike case however, this condition alone will not guarantee the existence of null COs. If we further require that the asymptotically flat spacetime should have a positive ADM mass, then clearly the corresponding metric will break the non-existence condition \refer{nstaticsol1}, but not necessarily the condition \refer{nstaticsol2}. Indeed, the metric \refer{theeg} is such an example: this metric is asymptotically flat and still satisfies Eq. \refer{nstaticsol2} and therefore has no null CO. Because for asymptotic flat SAS spacetimes with a positive ADM mass, Theorem \ref{asyflatexistence} guarantees the existence of timelike COs, the above example shows that for these spacetimes the existence of null COs has more stringent requirement.

In spite of our incapability of producing a general existence theorem for null COs, from physical considerations such as asymptotic flatness and positive ADM mass, we can still prove the following theorem that we believe might be useful for special spacetimes:\\
(Theorem \conc{bhnullext}) If there exist a finite range of $\rho\in[a,b)$ where $0<a<b$ such that the metric function $A(\rho=a)=0$ and $A(a<\rho<b)>0$, and $A(\rho\to\infty)$ does not diverge faster than $\rho^1$, then this metric allows null COs. \\
The proof of this is simple and given in Appendix \ref{proof1}. Since known BH spacetimes described by the metric \refer{metric2} usually has a radius $\rho_{\mbox{\scriptsize BH}}$ at which $A(\rho_{\mbox{\scriptsize BH}})=0$, this theorem might be especially useful to judge the existence of null COs around such BHs.

We can also consider the stability of null COs using the Lyapunov exponent. In this case, the Lyapunov exponents are
\bea
\lambda_\pm=\pm\sqrt{\frac{V^{\prime\prime}(\rho_*)}{2}}= \pm\sqrt{-\frac{2L^2A^{\prime\prime}(\rho_*)}{\rho_*^2D(\rho_*)}} \label{lambdanull}
\eea
We see that the only factor whose sign is not fixed and cannot be fixed by Eqs. \refer{nc1eq1} and \refer{nc1eq2}, is $A^{\prime\prime}(\rho_*)$. When $A^{\prime\prime}(\rho_*)$ is positive, zero or negative, the null CO will be stable, marginally stable or unstable respectively. Similar to the study of stability of timelike COs in section \ref{sectimestab}, we can equate $A^{\prime\prime}$ to some positive (or negative, or zero) definite arbitrary functions and solve for some general forms of $A(\rho)$ whose null COs will definitely be stable (or unstable, or have marginal stability). However, these general forms are too trivial to be used to judge stabilities of COs in known SAS spacetimes and therefore they are not shown here.

\section{Discussion\label{discuss}}

In our previous work \cite{Jia:2017nen}, the existence and stability of COs in static and spherically symmetric (SSS) spacetimes were studied and various theorems were obtained. Since SSS spacetime is also SAS, we first verify that the theorems obtained in the current work should also apply to the SSS case and therefore reproduce the theorems in Ref. \cite{Jia:2017nen} when the metric \refer{metric3} is reduced to the metric used in Ref. \cite{Jia:2017nen}:
\be \dd s^2=f(r)\dd t_{\mbox{\scriptsize S}}^2-\frac{1}{g(r)}\dd r^2-r^2(\dd\theta^2+\sin^2\theta\dd\phi_{\mbox{\scriptsize S}}^2), \ee
where $(t_{\mbox{\scriptsize S}},~r,~\theta,~\phi_{\mbox{\scriptsize S}})$ are the coordinates.
It is only necessary to consider the geodesics in the equatorial plane, defined by $\theta=\pi/2$ and therefore the SSS metric in this plane becomes
\be \dd s^2=f(r)\dd t_{\mbox{\scriptsize S}}^2-\frac{1}{g(r)}\dd r^2-r^2\dd\phi_{\mbox{\scriptsize S}}^2.\label{sss2d} \ee
It is then apparent the coordinate change transforming metric \refer{metric3} to metric \refer{sss2d} can be chosen as
\be
t_{\mbox{\scriptsize S}}=t, ~\phi_{\mbox{\scriptsize S}}=\phi, ~r^2=\frac{\rho^2}{A(\rho)},
\ee
and the metric functions should match as
\be f(r(\rho))=A(\rho),~g(r(\rho))=\frac{1}{D(\rho)}\frac{\dd \rho^2}{\dd r^2}. \ee
Under this transformation, the factor $\Phi(\rho)$ in the potential \refer{Vdef}, which becomes the starting point of the analysis in Eq. \refer{fpcon}, can be transformed to
\be
\Phi(\rho)\to\Phi(r)=\frac{1}{4r^2f(r)^2}\left(\frac{E^2}{f^2}-\frac{L^2}{r^2}-\epsilon\right). \label{transphi}\ee
Except an irrelevant factor, this is exact the potential $V(r)$ in Ref. \cite{Jia:2017nen}. Since all theorems in that work are then derived from this potential, they will be in agreement with the theorems obtained in current work.
This checks that the analysis and results for the equatorial motions in the SAS spacetime implies the results obtained in Ref. \cite{Jia:2017nen} for the SSS spacetime, as it should.

The original motivation of the work is to find COs for spacetimes whose metric functions are not analytically known. Usually in general relativistic models once the ansatz for the metric functions and matter content are chosen, their Einstein equations and Euler-Lagrangian equations for matter are easy to derive. Due to the non-linearity of these equations, analytical solutions are usually unavailable except in a few notable cases, and then numerical methods to different levels of complexity and difficulties, are used with the help of boundary conditions. In analyzing the COs in such metrics, we realized that the existence and stability of the COs can be determined without having to know the numerical solution. We now show as an example that the results in this work can be applied to metric in the EYMD model studied in Ref. \cite{Kleihaus:1997mn}. In this work the SAS ansatz for the metric in the equatorial plane is given by
\be \dd s^2= f(r)\dd t_{\mbox{\scriptsize EYMD}}^2-\frac{m(r)}{f(r)}\dd r ^2 -\frac{r^2l(r)}{f(r)}\dd\phi_{\mbox{\scriptsize EYMD}}^2, \ee
where $(t_{\mbox{\scriptsize EYMD}},~r,~\theta_{\mbox{\scriptsize EYMD}},~\phi_{\mbox{\scriptsize EYMD}})$ are the coordinates and $\theta_{\mbox{\scriptsize EYMD}}=\pi/2$ was set. We emphasise that the functions in this coefficient is analytically unknown and only some numerical solution are shown to exist.
It is not hard to verify that this metric is equivalent to metric \refer{metric3} if the following coordinate transforms are used
\be
t=t_{\mbox{\scriptsize EYMD}}, ~\phi=\phi_{\mbox{\scriptsize EYMD}},~\rho^2 =r^2l(r) \label{cotrans1} \ee
and metric functions are identified
\bea
&& A(\rho(r))=f(r),\label{functrans1}\\
&& D(\rho)^2=\frac{m(r)}{f(r)}\frac{\dd r^2}{\dd \rho^2}.
\eea
It is also known from Eq. (C10) of Ref. \cite{Kleihaus:1997mn} that the asymptotic expansion of the metric functions $f(r)$ and $l(r)$ are respectively
\bea
&& f(r)=1-\frac{c_f}{r}+\frac{c_f^2}{2r^2}+{\cal O}(r^{-3}), \label{fasy1}\\
&& l(r)=1-\frac{c_l}{r^2}+{\cal O}(r^{-3}), \label{lasy1}\eea
with $c_f>0$ and $c_l>0$ being constants. The relations \refer{cotrans1}, \refer{functrans1}, \refer{fasy1} and \refer{lasy1} imply that the asymptotic of $A(\rho)$ is
\be A(\rho)=1-\frac{c_f}{\rho} +\frac{c_f^2}{2\rho^2}+{\cal O}(\rho^{-3}) .\ee
This means that the metric function is asymptotically flat with a positive ADM mass. Therefore according to Theorem \ref{asyflatexistence} and \ref{flatplusunstab}, this spacetime will always allow timelike COs in its equatorial plane and the timelike COs at large $\rho$ are stable. This example shows the power of the results in this work.

Similar to our previous work \cite{Jia:2017nen}, we can also use the theorems obtained in this paper to study the existences and stabilities of the COs of some general SAS metrics. A total of four metrics from Ref. \cite{exactsolutionsofes} are examined in Table \ref{table1}. Their $A(\rho)$ components set at the equatorial plane are given in the first column, and their allowance of the timelike and null COs are listed in the second and third columns respectively. In column four we list for the metrics allowing COs whether they are asymptotically flat and if yes the sign of the ADM mass. It is clear that an asymptotic flat spacetime with a positive ADM mass always permits the existence of timelike COs, as we stated in Theorem \ref{asyflatexistence}, section \ref{sectimeasy}.

\end{multicols}

\pagestyle{empty}

\begin{landscape}

\begin{table*}[htp]
\caption{The $A(\rho)$ components of metrics of known SAS spacetimes from Ref. \cite{exactsolutionsofes}, their existence of timelike CO (Yes: Y, No: N), null CO (Yes: Y, No: N), asymptotic flatness (Flat: F, non-flat: N) and sign of corresponding ADM mass (semi-positive: $+$, semi-negative: $-$). See Ref. \cite{exactsolutionsofes}, Eqs. (20.4), (21.4), (21.7) and (21.10) for other components of the metrics. $a,~b,~m,~e$ and $\sigma$ in the metrics are real constants, $P_n(0)$ is the $n$-th order Legendre polynomial at zero. $n$ and $l$ are non-negative integers. \label{table1}}
\begin{center}
\begin{tabular}{l|l|l|l}
\hline\hline
$A(\rho)$&Timelike CO Exist.&Null CO Exist.&Asympt.\\
\hline
$\displaystyle\exp\left[2\sum_{n=0}^\infty a_{2n}P_{2n}(0)\rho^{-(2n+1)}\right]$
&Y ($a_{0,2,\cdots,2(l-1)}=0,~a_{2l}P_{2l}(0)<0$)       &Y ($a_{2l}P_{2l}(0)<0,~a_{\mbox{\scriptsize other}}=0$)                                                                                 &F, $+$ \\
&N ($a_{2l}P_{2l}(0)>0,~a_{\mbox{\scriptsize other}}=0$)&N ($a_{2l}P_{2l}(0)>0,~a_{\mbox{\scriptsize other}}=0$)                                                                                 &F, $-$ \\
&N ($a_{4l}\geq0,~a_{4l+2)}\leq0$)                      &N ($a_{4l}\geq0,~a_{4l+2)}\leq0$)  &F, $-$ \\
&Y ($a_{2n}=a^{2n+1},~a<0$)                             &N ($a_{2n}=a^{2n+1},~a<0$)         &F, $+$\\
&N ($a_{2n}=a^{2n+1},~a>0$)                             &N ($a_{2n}=a^{2n+1},~a>0$)         &F, $-$\\
&Y ($a_{2n}=na^{2n+1},~a>0$)                            &                                   &F, $+$ \\
\hline
$\frac{\rho^2}{\left(\sqrt{\rho^2+m^2-e^2}+m\right)^2},~\rho>e$
&Y ($m>0$)                                              &Y ($m>2\sqrt{2}|e|/3$)             &F, $+$\\
&                                                       &N ($0<m<2\sqrt{2}|e|/3$)           &F, $+$ \\
&N ($m<0$)                                              &N ($m<0$)                          &F, $-$ \\
\hline
$\left[\frac{\rho^2+(1+a b)\sigma^2}{\left(a\sigma+\sqrt{\rho^2+\sigma^2}\right)\left(-b\sigma+\sqrt{\rho^2+\sigma^2}\right)}\right]^2$
&Y ($a>b,~\sigma>0$)                                    &                                   &F, $-$ \\
\hline
$\frac{x-1}{x+1}\frac{\left[x^2+\alpha^2(x^2-1)\right]^2+4\alpha^2x^2}{\left[x^2+\alpha^2(x^2-1)\right]^2}, ~x=\frac{\sqrt{\rho^2+\sigma^2}}{\sigma}$
&Y ($\sigma>0$)                                         &                                   &F, $-$\\
\hline\hline
\end{tabular}
\end{center}
\end{table*}

\end{landscape}

\begin{multicols}{2}
For the extension of the current work, two possible directions are of special interests. The first is to extend  the analysis for SAS spacetimes to stationary and axisymmetric spacetimes because many axisymmetric and physically important spacetimes are stationary but not static. An analysis for the COs in stationary spacetimes therefore will be more useful in the perspective of applications. In the stationary case, since $B(\rho)$ in \refer{metric2} will be nonzero, the equations analogous to Eq. \refer{eqmat} will contains two arbitrary functions $A(\rho)$ and $B(\rho)$ rather than one $A(\rho)$ and a cross-terms proportional to $EL$ will appear. Even though these make the analysis more complicated, some preliminary work shows that the (non-)existence condition might be still solvable.

Another direction of extension is to consider the stability issues of non-radical perturbations in general stationary and axisymmetric spacetimes due to its relevance to the dynamics of a finitely thick disk in the accretion phase of the (binary) rotating systems. The zenithal stability issue in particular axisymmetric spacetimes has been studied recently in Ref. \cite{Ono:2016lql}  and an analysis using Lyapunov exponent method for general metrics might reveal more general stability conditions on metric functions.

\appendix

\section{Sufficient conditions for (in-)stabilities of timelike CO\texorpdfstring{\MakeLowercase{s}}{} \label{appda}}

When Eq. \refer{v2cb} is replaced  by
\be
h(\rho)=\chi\rho^2A^{\prime3}~(\chi\neq0), \label{v2cbc1}
\ee
an implicit solution can be obtained
\be \frac {\sqrt {1+8\chi}A^{\frac{3}{2}+\frac{\sqrt {1+8\chi}}{2}}}{2\rho^2} -c_2 A^{\sqrt {1+8\chi}}+c_1=0. \ee
In particular, when $\chi=1$ or $\chi=-2/25$, two explicit solution can be found
\bea
\chi=1, &&A (\rho )=\left({\frac {2c_1\rho^2}{2c_2\rho^2-3}}\right)^{1/3}\\
\chi=-\frac{2}{25}, && A(\rho)=\left(\frac{1}{3} x+\frac{10}{3}x^{-1}\right)^{5/3},\\
x&=&\left(5\sqrt {81c_1^2\rho^4-40c_2^3\rho^6}-45c_1\rho^2\right)^{1/3}. \nonumber\eea
The former will only have stable COs (if any) and the later will only have unstable COs (if any) and none of them will allow MSCOs.

When Eq. \refer{v2cb} is replaced  by
\be
h(\rho)=\chi A^2A^\prime ~(\chi\neq0),\label{v2cbc2}
\ee
an implicit solution can be obtained
\be  ( 2-\chi )\left(c_1- c_2 A^{\sqrt {1+4\chi }}\right) \rho^5 + \sqrt {1+4\chi }A^{\frac{3}{2}+\frac{\sqrt {1+4\chi }}{2}}\rho^{\chi +3}=0.
 \ee
In particular, when $\chi=-4/25$, an explicit solution can be found
\bea
&&A(\rho)=\left[\left( y/5+6c_2\rho^{\frac {4}{25}}
y^{-1} \right) \rho\right]^{5/3},\\
&& y=\left( 15\sqrt {225c_1^2- 120c_2^3 \rho^{\frac {54}{25}}}-225c_1 \right)^{1/3} \rho^{-\frac {7}{25}}. \nonumber\eea
This solution will only have unstable COs (if any) and no MSCOs.

When Eq. \refer{v2cb} is replaced  by
\be
h(\rho)=\chi \rho AA^{\prime2} ~(\chi\neq0), \label{v2cbc3} \ee
an implicit solution can be obtained
\be
c_1- c_2 A^{\sqrt {\chi ^2+6\chi +1}}+ \frac{\sqrt{\chi ^2+6\chi +1} A^{\frac{\chi}{2}+ \frac{3}{2}+\frac{\sqrt{\chi ^2+6\chi +1}}{2}}} {2\rho^2}=0.  \ee
In particular, when $\chi=-3+5\sqrt{3}/3$, an explicit solution can be found
\bea
&&A(\rho)=\left(\frac{u}{3}+2\sqrt{3}c_2\rho^2 u^{-1}\right)^{\sqrt{3}},\\
&&u=\left(9\sqrt{27{c_1}^2\rho^4-8\sqrt{3}c_2^3\rho^6} -27\sqrt{3}c_1\rho^2\right)^{1/3}. \nonumber\eea
This solution will only have unstable COs (if any) and no MSCOs.

One can also replace Eq. \refer{v2cb} by
\be h(\rho)=\chi \rho A^2A^{\prime\prime} ~(\chi\neq0).\label{v2cbc4}\ee
Although the sign of $A^{\prime\prime}$ and consequently that of the right hand side cannot be fixed before solving $A(\rho)$, this transformation do allow the equation to be solvable and then a determination of $A^{\prime\prime}$'s sign. The implicit solution to this equation is given by
\be
( \chi +2 ) c_2 A^{-\frac {1+\chi }{\chi -1}}+ ( 1+\chi  ) A^{-\frac {\chi +2}{\chi -1}}\rho^{\frac {\chi +2}{\chi -1}} +c_1=0. \label{c4sol}
 \ee
In particular, when $\chi=-3$, two explicit solution can be found
\bea && A(\rho)=\left[\left(\rho^{1/4}+\sqrt {c_1c_2+\sqrt {\rho}}\right)/c_1\right]^4,\\
&& A(\rho)=\left[\left(-\rho^{1/4}+\sqrt {c_1c_2+\sqrt {\rho}}\right)/c_1\right]^4
, \eea
where in both cases $c_1c_2>0$. For the former solution, since $A^{\prime\prime}<0$ it will only have unstable COs (if any) and no MSCOs. For the later one, $A^{\prime\prime}>0$ and it will only have stable COs (if any) and no MSCOs. When $\chi=-1/2$, or $\chi=-5/2$, Eq. \refer{c4sol} can also generate some explicit solutions of $A(\rho)$. However, since sign of their second derivatives cannot be determined, and we will not list them here.

\section{Proof of Theorem \ref{bhnullext}\label{proof1}}

To prove the theorem, we define a $\mu(\rho)$ first
\be \mu(\rho)=A(\rho)/\rho \ee
such that
\be \mu^\prime(\rho)=\sqb{A^\prime(\rho)\rho-A(\rho)}/\rho^2 .\ee
Then using Eq. \refer{nc1eq2} it is seen that the existence of a null CO is equivalent to the existence of a point such that $\mu^\prime(\rho)=0$.

Now since $A(\rho=a)=0$,  we have
\be \mu^\prime(\rho=a)=A^\prime(\rho=a)/a^2. \ee
Since $A(\rho)>0$ for at least the neighborhood $(a,~b)$ of $\rho$, we must have $A^\prime(\rho=a)\geq0$. Therefore
\be \mu^\prime(\rho=a)\geq0 .\ee
If the equal sign is true, then the CO exists.
If it is not true, we further show by contradiction that $\mu^\prime(\rho)$ cannot be positive definite for $\rho\in[a,~\infty)$ and therefore due to the continuity of $\mu^\prime(\rho)$ there must exist a point $\rho_*$ satisfying $\mu^\prime(\rho_*)=0$, which is again a CO.
Suppose $\mu^\prime(\rho)>0$ for all $\rho\in[a,~\infty)$, then clearly  $A(\rho)=\mu(\rho)\rho$ will diverge faster than $\rho^1$, which contradicts the assumption.

\bigskip

\bibliographystyle{apsrev4-1}
\bibliography{asfp_ref}

\begin{thebibliography}{32}%
\makeatletter
\providecommand \@ifxundefined [1]{%
 \@ifx{#1\undefined}
}%
\providecommand \@ifnum [1]{%
 \ifnum #1\expandafter \@firstoftwo
 \else \expandafter \@secondoftwo
 \fi
}%
\providecommand \@ifx [1]{%
 \ifx #1\expandafter \@firstoftwo
 \else \expandafter \@secondoftwo
 \fi
}%
\providecommand \natexlab [1]{#1}%
\providecommand \enquote  [1]{``#1''}%
\providecommand \bibnamefont  [1]{#1}%
\providecommand \bibfnamefont [1]{#1}%
\providecommand \citenamefont [1]{#1}%
\providecommand \href@noop [0]{\@secondoftwo}%
\providecommand \href [0]{\begingroup \@sanitize@url \@href}%
\providecommand \@href[1]{\@@startlink{#1}\@@href}%
\providecommand \@@href[1]{\endgroup#1\@@endlink}%
\providecommand \@sanitize@url [0]{\catcode `\\12\catcode `\$12\catcode
  `\&12\catcode `\#12\catcode `\^12\catcode `\_12\catcode `\%12\relax}%
\providecommand \@@startlink[1]{}%
\providecommand \@@endlink[0]{}%
\providecommand \url  [0]{\begingroup\@sanitize@url \@url }%
\providecommand \@url [1]{\endgroup\@href {#1}{\urlprefix }}%
\providecommand \urlprefix  [0]{URL }%
\providecommand \Eprint [0]{\href }%
\providecommand \doibase [0]{http://dx.doi.org/}%
\providecommand \selectlanguage [0]{\@gobble}%
\providecommand \bibinfo  [0]{\@secondoftwo}%
\providecommand \bibfield  [0]{\@secondoftwo}%
\providecommand \translation [1]{[#1]}%
\providecommand \BibitemOpen [0]{}%
\providecommand \bibitemStop [0]{}%
\providecommand \bibitemNoStop [0]{.\EOS\space}%
\providecommand \EOS [0]{\spacefactor3000\relax}%
\providecommand \BibitemShut  [1]{\csname bibitem#1\endcsname}%
\let\auto@bib@innerbib\@empty
\bibitem [{\citenamefont {Del~Zanna}\ \emph {et~al.}(2004)\citenamefont
  {Del~Zanna}, \citenamefont {Amato},\ and\ \citenamefont
  {Bucciantini}}]{DelZanna:2004aq}%
  \BibitemOpen
  \bibfield  {author} {\bibinfo {author} {\bibfnamefont {L.}~\bibnamefont
  {Del~Zanna}}, \bibinfo {author} {\bibfnamefont {E.}~\bibnamefont {Amato}}, \
  and\ \bibinfo {author} {\bibfnamefont {N.}~\bibnamefont {Bucciantini}},\
  }\href {\doibase doi:10.1051/0004-6361:20035936} {\bibfield  {journal}
  {\bibinfo  {journal} {Astron. Astrophys.}\ }\textbf {\bibinfo {volume}
  {421}},\ \bibinfo {pages} {1063} (\bibinfo {year} {2004})},\ \bibinfo {note}
  {doi:10.1051/0004-6361:20035936 [astro-ph/0404355]}\BibitemShut {NoStop}%
\bibitem [{\citenamefont {et~al.}(2002)}]{Wang:2002rx}%
  \BibitemOpen
  \bibfield  {author} {\bibinfo {author} {\bibfnamefont {L.~W.}\ \bibnamefont
  {et~al.}},\ }\href {\doibase doi:10.1086/342824} {\bibfield  {journal}
  {\bibinfo  {journal} {Astrophys. J.}\ }\textbf {\bibinfo {volume} {579}},\
  \bibinfo {pages} {671} (\bibinfo {year} {2002})},\ \bibinfo {note}
  {doi:10.1086/342824 [astro-ph/0205337]}\BibitemShut {NoStop}%
\bibitem [{\citenamefont {Radu}(2002)}]{Radu:2001ij}%
  \BibitemOpen
  \bibfield  {author} {\bibinfo {author} {\bibfnamefont {E.}~\bibnamefont
  {Radu}},\ }\href {\doibase doi:10.1103/PhysRevD.65.044005} {\bibfield
  {journal} {\bibinfo  {journal} {Phys. Rev. D}\ }\textbf {\bibinfo {volume}
  {65}},\ \bibinfo {pages} {044005} (\bibinfo {year} {2002})},\ \bibinfo {note}
  {doi:10.1103/PhysRevD.65.044005 [gr-qc/0109015]}\BibitemShut {NoStop}%
\bibitem [{\citenamefont {Hartmann}\ \emph {et~al.}(2001)\citenamefont
  {Hartmann}, \citenamefont {Kleihaus},\ and\ \citenamefont
  {Kunz}}]{Hartmann:2001ic}%
  \BibitemOpen
  \bibfield  {author} {\bibinfo {author} {\bibfnamefont {B.}~\bibnamefont
  {Hartmann}}, \bibinfo {author} {\bibfnamefont {B.}~\bibnamefont {Kleihaus}},
  \ and\ \bibinfo {author} {\bibfnamefont {J.}~\bibnamefont {Kunz}},\ }\href
  {\doibase doi:10.1103/PhysRevD.65.024027} {\bibfield  {journal} {\bibinfo
  {journal} {Phys. Rev. D}\ }\textbf {\bibinfo {volume} {65}},\ \bibinfo
  {pages} {024027} (\bibinfo {year} {2001})},\ \bibinfo {note}
  {doi:10.1103/PhysRevD.65.024027 [hep-th/0108129]}\BibitemShut {NoStop}%
\bibitem [{\citenamefont {Kleihaus}\ and\ \citenamefont
  {Kunz}(1997)}]{Kleihaus:1996vi}%
  \BibitemOpen
  \bibfield  {author} {\bibinfo {author} {\bibfnamefont {B.}~\bibnamefont
  {Kleihaus}}\ and\ \bibinfo {author} {\bibfnamefont {J.}~\bibnamefont
  {Kunz}},\ }\href {\doibase doi:10.1103/PhysRevLett.78.2527} {\bibfield
  {journal} {\bibinfo  {journal} {Phys. Rev. Lett.}\ }\textbf {\bibinfo
  {volume} {78}},\ \bibinfo {pages} {2527} (\bibinfo {year} {1997})},\ \bibinfo
  {note} {doi:10.1103/PhysRevLett.78.2527 [hep-th/9612101]}\BibitemShut
  {NoStop}%
\bibitem [{\citenamefont {Kleihaus}\ and\ \citenamefont
  {Kunz}(1998{\natexlab{a}})}]{Kleihaus:1997ws}%
  \BibitemOpen
  \bibfield  {author} {\bibinfo {author} {\bibfnamefont {B.}~\bibnamefont
  {Kleihaus}}\ and\ \bibinfo {author} {\bibfnamefont {J.}~\bibnamefont
  {Kunz}},\ }\href {\doibase doi:10.1103/PhysRevD.57.6138} {\bibfield
  {journal} {\bibinfo  {journal} {Phys. Rev. D}\ }\textbf {\bibinfo {volume}
  {57}},\ \bibinfo {pages} {6138} (\bibinfo {year} {1998}{\natexlab{a}})},\
  \bibinfo {note} {doi:10.1103/PhysRevD.57.6138 [gr-qc/9712086]}\BibitemShut
  {NoStop}%
\bibitem [{\citenamefont {Kleihaus}\ and\ \citenamefont
  {Kunz}(1998{\natexlab{b}})}]{Kleihaus:1997mn}%
  \BibitemOpen
  \bibfield  {author} {\bibinfo {author} {\bibfnamefont {B.}~\bibnamefont
  {Kleihaus}}\ and\ \bibinfo {author} {\bibfnamefont {J.}~\bibnamefont
  {Kunz}},\ }\href {\doibase doi:10.1103/PhysRevD.57.834} {\bibfield  {journal}
  {\bibinfo  {journal} {Phys. Rev. D}\ }\textbf {\bibinfo {volume} {57}},\
  \bibinfo {pages} {834} (\bibinfo {year} {1998}{\natexlab{b}})},\ \bibinfo
  {note} {doi:10.1103/PhysRevD.57.834 [gr-qc/9707045]}\BibitemShut {NoStop}%
\bibitem [{\citenamefont {Capozziello}\ \emph {et~al.}(2010)\citenamefont
  {Capozziello}, \citenamefont {De~Laurentis},\ and\ \citenamefont
  {Stabile}}]{Capozziello:2009jg}%
  \BibitemOpen
  \bibfield  {author} {\bibinfo {author} {\bibfnamefont {S.}~\bibnamefont
  {Capozziello}}, \bibinfo {author} {\bibfnamefont {M.}~\bibnamefont
  {De~Laurentis}}, \ and\ \bibinfo {author} {\bibfnamefont {A.}~\bibnamefont
  {Stabile}},\ }\href {\doibase doi:10.1088/0264-9381/27/16/165008} {\bibfield
  {journal} {\bibinfo  {journal} {Class. Quant. Grav.}\ }\textbf {\bibinfo
  {volume} {27}},\ \bibinfo {pages} {165008} (\bibinfo {year} {2010})},\
  \bibinfo {note} {doi:10.1088/0264-9381/27/16/165008 [arXiv:0912.5286
  [gr-qc]]}\BibitemShut {NoStop}%
\bibitem [{\citenamefont {Kuhfittig}(2003)}]{Kuhfittig:2003wr}%
  \BibitemOpen
  \bibfield  {author} {\bibinfo {author} {\bibfnamefont {P.~K.}\ \bibnamefont
  {Kuhfittig}},\ }\href {\doibase doi:10.1103/PhysRevD.67.064015} {\bibfield
  {journal} {\bibinfo  {journal} {Phys. Rev. D}\ }\textbf {\bibinfo {volume}
  {67}},\ \bibinfo {pages} {064015} (\bibinfo {year} {2003})},\ \bibinfo {note}
  {doi:10.1103/PhysRevD.67.064015 [gr-qc/0401028]}\BibitemShut {NoStop}%
\bibitem [{\citenamefont {Reddy}\ \emph {et~al.}(2006)\citenamefont {Reddy},
  \citenamefont {Naidu},\ and\ \citenamefont {Rao}}]{Reddy:2006np}%
  \BibitemOpen
  \bibfield  {author} {\bibinfo {author} {\bibfnamefont {D.~R.}\ \bibnamefont
  {Reddy}}, \bibinfo {author} {\bibfnamefont {R.}~\bibnamefont {Naidu}}, \ and\
  \bibinfo {author} {\bibfnamefont {V.~U.}\ \bibnamefont {Rao}},\ }\href
  {\doibase doi:10.1007/s10509-006-9169-x} {\bibfield  {journal} {\bibinfo
  {journal} {Astrophys. Space Sci.}\ }\textbf {\bibinfo {volume} {306}},\
  \bibinfo {pages} {185} (\bibinfo {year} {2006})}\BibitemShut {NoStop}%
\bibitem [{\citenamefont {Reddy}\ and\ \citenamefont
  {Rao}(2006)}]{Reddy:2006dz}%
  \BibitemOpen
  \bibfield  {author} {\bibinfo {author} {\bibfnamefont {D.~R.}\ \bibnamefont
  {Reddy}}\ and\ \bibinfo {author} {\bibfnamefont {M.~V.~S.}\ \bibnamefont
  {Rao}},\ }\href {\doibase doi:10.1007/s10509-005-9022-7} {\bibfield
  {journal} {\bibinfo  {journal} {Astrophys. Space Sci.}\ }\textbf {\bibinfo
  {volume} {302}},\ \bibinfo {pages} {157} (\bibinfo {year}
  {2006})}\BibitemShut {NoStop}%
\bibitem [{\citenamefont {Vlachynsky}\ \emph {et~al.}(1996)\citenamefont
  {Vlachynsky}, \citenamefont {Tresguerres}, \citenamefont {Obukhov},\ and\
  \citenamefont {Hehl}}]{Vlachynsky:1996zh}%
  \BibitemOpen
  \bibfield  {author} {\bibinfo {author} {\bibfnamefont {E.}~\bibnamefont
  {Vlachynsky}}, \bibinfo {author} {\bibfnamefont {R.}~\bibnamefont
  {Tresguerres}}, \bibinfo {author} {\bibfnamefont {Y.~N.}\ \bibnamefont
  {Obukhov}}, \ and\ \bibinfo {author} {\bibfnamefont {F.}~\bibnamefont
  {Hehl}},\ }\href {\doibase doi:10.1088/0264-9381/13/12/016} {\bibfield
  {journal} {\bibinfo  {journal} {Class. Quant. Grav.}\ }\textbf {\bibinfo
  {volume} {13}},\ \bibinfo {pages} {3253} (\bibinfo {year} {1996})},\ \bibinfo
  {note} {doi:10.1088/0264-9381/13/12/016 [gr-qc/9604035]}\BibitemShut
  {NoStop}%
\bibitem [{\citenamefont {Abbott}\ \emph
  {et~al.}(2016{\natexlab{a}})\citenamefont {Abbott}, \citenamefont {Abbott},
  \citenamefont {Abbott}, \citenamefont {Abernathy}, \citenamefont {Acernese},
  \citenamefont {Ackley}, \citenamefont {Adams}, \citenamefont {Adams},
  \citenamefont {Addesso}, \citenamefont {Adhikari} \emph
  {et~al.}}]{Abbott:2016blz}%
  \BibitemOpen
  \bibfield  {author} {\bibinfo {author} {\bibfnamefont {B.}~\bibnamefont
  {Abbott}}, \bibinfo {author} {\bibfnamefont {R.}~\bibnamefont {Abbott}},
  \bibinfo {author} {\bibfnamefont {T.}~\bibnamefont {Abbott}}, \bibinfo
  {author} {\bibfnamefont {M.}~\bibnamefont {Abernathy}}, \bibinfo {author}
  {\bibfnamefont {F.}~\bibnamefont {Acernese}}, \bibinfo {author}
  {\bibfnamefont {K.}~\bibnamefont {Ackley}}, \bibinfo {author} {\bibfnamefont
  {C.}~\bibnamefont {Adams}}, \bibinfo {author} {\bibfnamefont
  {T.}~\bibnamefont {Adams}}, \bibinfo {author} {\bibfnamefont
  {P.}~\bibnamefont {Addesso}}, \bibinfo {author} {\bibfnamefont
  {R.}~\bibnamefont {Adhikari}},  \emph {et~al.},\ }\href {\doibase
  doi:10.1103/PhysRevLett.116.061102} {\bibfield  {journal} {\bibinfo
  {journal} {Phys. Rev. Lett.}\ }\textbf {\bibinfo {volume} {116}},\ \bibinfo
  {pages} {061102} (\bibinfo {year} {2016}{\natexlab{a}})},\ \bibinfo {note}
  {doi:10.1103/PhysRevLett.116.061102 [arXiv:1602.03837 [gr-qc]]}\BibitemShut
  {NoStop}%
\bibitem [{\citenamefont {Abbott}\ \emph
  {et~al.}(2016{\natexlab{b}})\citenamefont {Abbott}, \citenamefont {Abbott},
  \citenamefont {Abbott}, \citenamefont {Abernathy}, \citenamefont {Acernese},
  \citenamefont {Ackley}, \citenamefont {Adams}, \citenamefont {Adams},
  \citenamefont {Addesso}, \citenamefont {Adhikari} \emph
  {et~al.}}]{Abbott:2016nmj}%
  \BibitemOpen
  \bibfield  {author} {\bibinfo {author} {\bibfnamefont {B.}~\bibnamefont
  {Abbott}}, \bibinfo {author} {\bibfnamefont {R.}~\bibnamefont {Abbott}},
  \bibinfo {author} {\bibfnamefont {T.}~\bibnamefont {Abbott}}, \bibinfo
  {author} {\bibfnamefont {M.}~\bibnamefont {Abernathy}}, \bibinfo {author}
  {\bibfnamefont {F.}~\bibnamefont {Acernese}}, \bibinfo {author}
  {\bibfnamefont {K.}~\bibnamefont {Ackley}}, \bibinfo {author} {\bibfnamefont
  {C.}~\bibnamefont {Adams}}, \bibinfo {author} {\bibfnamefont
  {T.}~\bibnamefont {Adams}}, \bibinfo {author} {\bibfnamefont
  {P.}~\bibnamefont {Addesso}}, \bibinfo {author} {\bibfnamefont
  {R.}~\bibnamefont {Adhikari}},  \emph {et~al.},\ }\href {\doibase
  doi:10.1103/PhysRevLett.116.241103} {\bibfield  {journal} {\bibinfo
  {journal} {Phys. Rev. Lett.}\ }\textbf {\bibinfo {volume} {116}},\ \bibinfo
  {pages} {241103} (\bibinfo {year} {2016}{\natexlab{b}})},\ \bibinfo {note}
  {doi:10.1103/PhysRevLett.116.241103 [arXiv:1606.04855 [gr-qc]]}\BibitemShut
  {NoStop}%
\bibitem [{\citenamefont {Hackmann}\ and\ \citenamefont
  {L{\"a}mmerzahl}(2012)}]{Hackmann:2011wp}%
  \BibitemOpen
  \bibfield  {author} {\bibinfo {author} {\bibfnamefont {E.}~\bibnamefont
  {Hackmann}}\ and\ \bibinfo {author} {\bibfnamefont {C.}~\bibnamefont
  {L{\"a}mmerzahl}},\ }\href {\doibase doi:10.1103/PhysRevD.85.044049}
  {\bibfield  {journal} {\bibinfo  {journal} {Phys. Rev. D}\ }\textbf {\bibinfo
  {volume} {85}},\ \bibinfo {pages} {044049} (\bibinfo {year} {2012})},\
  \bibinfo {note} {doi:10.1103/PhysRevD.85.044049 [arXiv:1107.5250
  [gr-qc]]}\BibitemShut {NoStop}%
\bibitem [{\citenamefont {Sanabria-Gomez}\ \emph {et~al.}(2010)\citenamefont
  {Sanabria-Gomez}, \citenamefont {Hernandez-Pastora},\ and\ \citenamefont
  {Dubeibe}}]{SanabriaGomez:2010vb}%
  \BibitemOpen
  \bibfield  {author} {\bibinfo {author} {\bibfnamefont {J.~D.}\ \bibnamefont
  {Sanabria-Gomez}}, \bibinfo {author} {\bibfnamefont {J.~L.}\ \bibnamefont
  {Hernandez-Pastora}}, \ and\ \bibinfo {author} {\bibfnamefont {F.~L.}\
  \bibnamefont {Dubeibe}},\ }\href {\doibase 10.1103/PhysRevD.82.124014}
  {\bibfield  {journal} {\bibinfo  {journal} {Phys. Rev.}\ }\textbf {\bibinfo
  {volume} {D82}},\ \bibinfo {pages} {124014} (\bibinfo {year} {2010})},\
  \Eprint {http://arxiv.org/abs/1009.0320} {arXiv:1009.0320 [gr-qc]}
  \BibitemShut {NoStop}%
\bibitem [{\citenamefont {Thomas}\ \emph {et~al.}(2004)\citenamefont {Thomas},
  \citenamefont {Saglia}, \citenamefont {Bender}, \citenamefont {Thomas},
  \citenamefont {Gebhardt}, \citenamefont {Magorrian},\ and\ \citenamefont
  {Richstone}}]{Thomas:2004rz}%
  \BibitemOpen
  \bibfield  {author} {\bibinfo {author} {\bibfnamefont {J.}~\bibnamefont
  {Thomas}}, \bibinfo {author} {\bibfnamefont {R.}~\bibnamefont {Saglia}},
  \bibinfo {author} {\bibfnamefont {R.}~\bibnamefont {Bender}}, \bibinfo
  {author} {\bibfnamefont {D.}~\bibnamefont {Thomas}}, \bibinfo {author}
  {\bibfnamefont {K.}~\bibnamefont {Gebhardt}}, \bibinfo {author}
  {\bibfnamefont {J.}~\bibnamefont {Magorrian}}, \ and\ \bibinfo {author}
  {\bibfnamefont {D.}~\bibnamefont {Richstone}},\ }\href {\doibase
  doi:10.1103/PhysRevD.94.064042} {\bibfield  {journal} {\bibinfo  {journal}
  {Mon. Not. Roy. Aston. Soc.}\ }\textbf {\bibinfo {volume} {353}},\ \bibinfo
  {pages} {391} (\bibinfo {year} {2004})},\ \bibinfo {note}
  {doi:10.1103/PhysRevD.94.064042 [arXiv:1605.05816 [gr-qc]]}\BibitemShut
  {NoStop}%
\bibitem [{\citenamefont {Shibata}\ and\ \citenamefont
  {Sasaki}(1998)}]{Shibata:1998xw}%
  \BibitemOpen
  \bibfield  {author} {\bibinfo {author} {\bibfnamefont {M.}~\bibnamefont
  {Shibata}}\ and\ \bibinfo {author} {\bibfnamefont {M.}~\bibnamefont
  {Sasaki}},\ }\href {\doibase doi:10.1103/PhysRevD.58.104011} {\bibfield
  {journal} {\bibinfo  {journal} {Phys. Rev. D}\ }\textbf {\bibinfo {volume}
  {58}},\ \bibinfo {pages} {104011} (\bibinfo {year} {1998})},\ \bibinfo {note}
  {doi:10.1103/PhysRevD.58.104011 [gr-qc/9807046]}\BibitemShut {NoStop}%
\bibitem [{\citenamefont {Donati}\ \emph {et~al.}(2005)\citenamefont {Donati},
  \citenamefont {Paletou}, \citenamefont {Bouvier},\ and\ \citenamefont
  {Ferreira}}]{Donati:2005tw}%
  \BibitemOpen
  \bibfield  {author} {\bibinfo {author} {\bibfnamefont {J.-F.}\ \bibnamefont
  {Donati}}, \bibinfo {author} {\bibfnamefont {F.}~\bibnamefont {Paletou}},
  \bibinfo {author} {\bibfnamefont {J.}~\bibnamefont {Bouvier}}, \ and\
  \bibinfo {author} {\bibfnamefont {J.}~\bibnamefont {Ferreira}},\ }\href
  {\doibase doi:10.1038/nature04253} {\bibfield  {journal} {\bibinfo  {journal}
  {Nature}\ }\textbf {\bibinfo {volume} {438}},\ \bibinfo {pages} {466}
  (\bibinfo {year} {2005})},\ \bibinfo {note} {doi:10.1038/nature04253
  [astro-ph/0511695]}\BibitemShut {NoStop}%
\bibitem [{\citenamefont {Abramowicz}\ \emph {et~al.}(2010)\citenamefont
  {Abramowicz}, \citenamefont {Jaroszy{\'n}ski}, \citenamefont {Kato},
  \citenamefont {Lasota}, \citenamefont {R{\'o}{\.z}a{\'n}ska},\ and\
  \citenamefont {S{\k{a}}dowski}}]{Abramowicz:2010nk}%
  \BibitemOpen
  \bibfield  {author} {\bibinfo {author} {\bibfnamefont {M.~A.}\ \bibnamefont
  {Abramowicz}}, \bibinfo {author} {\bibfnamefont {M.}~\bibnamefont
  {Jaroszy{\'n}ski}}, \bibinfo {author} {\bibfnamefont {S.}~\bibnamefont
  {Kato}}, \bibinfo {author} {\bibfnamefont {J.-P.}\ \bibnamefont {Lasota}},
  \bibinfo {author} {\bibfnamefont {A.}~\bibnamefont {R{\'o}{\.z}a{\'n}ska}}, \
  and\ \bibinfo {author} {\bibfnamefont {A.}~\bibnamefont {S{\k{a}}dowski}},\
  }\href {\doibase doi:10.1051/0004-6361/201014467} {\bibfield  {journal}
  {\bibinfo  {journal} {Astron. Astrophys.}\ }\textbf {\bibinfo {volume}
  {521}},\ \bibinfo {pages} {A15} (\bibinfo {year} {2010})},\ \bibinfo {note}
  {doi:10.1051/0004-6361/201014467 [arXiv:1003.3887 [astro-ph.HE]]}\BibitemShut
  {NoStop}%
\bibitem [{\citenamefont {Jia}\ \emph {et~al.}(2017)\citenamefont {Jia},
  \citenamefont {Liu}, \citenamefont {Liu}, \citenamefont {Mo}, \citenamefont
  {Pang}, \citenamefont {Wang},\ and\ \citenamefont {Yang}}]{Jia:2017nen}%
  \BibitemOpen
  \bibfield  {author} {\bibinfo {author} {\bibfnamefont {J.}~\bibnamefont
  {Jia}}, \bibinfo {author} {\bibfnamefont {J.}~\bibnamefont {Liu}}, \bibinfo
  {author} {\bibfnamefont {X.}~\bibnamefont {Liu}}, \bibinfo {author}
  {\bibfnamefont {Z.}~\bibnamefont {Mo}}, \bibinfo {author} {\bibfnamefont
  {X.}~\bibnamefont {Pang}}, \bibinfo {author} {\bibfnamefont {Y.}~\bibnamefont
  {Wang}}, \ and\ \bibinfo {author} {\bibfnamefont {N.}~\bibnamefont {Yang}},\
  }\href@noop {} {\bibfield  {journal} {\bibinfo  {journal} {arXiv:1702.05889,
  accepted by Gen. Rel. \& Grav.}\ } (\bibinfo {year} {2017})}\BibitemShut
  {NoStop}%
\bibitem [{\citenamefont {Letelier}(2003)}]{Letelier:2003ea}%
  \BibitemOpen
  \bibfield  {author} {\bibinfo {author} {\bibfnamefont {P.~S.}\ \bibnamefont
  {Letelier}},\ }\href {\doibase doi:10.1103/PhysRevD.68.104002} {\bibfield
  {journal} {\bibinfo  {journal} {Phys. Rev. D}\ }\textbf {\bibinfo {volume}
  {68}},\ \bibinfo {pages} {104002} (\bibinfo {year} {2003})},\ \bibinfo {note}
  {doi:10.1103/PhysRevD.68.104002 [gr-qc/0309033]}\BibitemShut {NoStop}%
\bibitem [{\citenamefont {L{\'o}pez-Suspes}\ and\ \citenamefont
  {Gonz{\'a}lez}(2014)}]{Gonzalez:2011fb}%
  \BibitemOpen
  \bibfield  {author} {\bibinfo {author} {\bibfnamefont {F.}~\bibnamefont
  {L{\'o}pez-Suspes}}\ and\ \bibinfo {author} {\bibfnamefont {G.~A.}\
  \bibnamefont {Gonz{\'a}lez}},\ }\href {\doibase
  doi:10.1007/s13538-014-0216-8} {\bibfield  {journal} {\bibinfo  {journal}
  {Braz. J. Phys.}\ }\textbf {\bibinfo {volume} {44}},\ \bibinfo {pages} {385}
  (\bibinfo {year} {2014})},\ \bibinfo {note} {doi:10.1007/s13538-014-0216-8
  [arXiv:1104.0346 [gr-qc]]}\BibitemShut {NoStop}%
\bibitem [{\citenamefont {Dolan}\ and\ \citenamefont
  {Shipley}(2016)}]{Dolan:2016bxj}%
  \BibitemOpen
  \bibfield  {author} {\bibinfo {author} {\bibfnamefont {S.~R.}\ \bibnamefont
  {Dolan}}\ and\ \bibinfo {author} {\bibfnamefont {J.~O.}\ \bibnamefont
  {Shipley}},\ }\href {\doibase doi:10.1103/PhysRevD.94.044038} {\bibfield
  {journal} {\bibinfo  {journal} {Phys. Rev. D}\ }\textbf {\bibinfo {volume}
  {94}},\ \bibinfo {pages} {044038} (\bibinfo {year} {2016})},\ \bibinfo {note}
  {doi:10.1103/PhysRevD.94.044038 [arXiv:1104.0346 [gr-qc]]}\BibitemShut
  {NoStop}%
\bibitem [{\citenamefont {Beheshti}\ and\ \citenamefont
  {Gasper{\'\i}n}(2016)}]{Beheshti:2015bak}%
  \BibitemOpen
  \bibfield  {author} {\bibinfo {author} {\bibfnamefont {S.}~\bibnamefont
  {Beheshti}}\ and\ \bibinfo {author} {\bibfnamefont {E.}~\bibnamefont
  {Gasper{\'\i}n}},\ }\href {\doibase doi:10.1103/PhysRevD.94.024015}
  {\bibfield  {journal} {\bibinfo  {journal} {Phys. Rev. D}\ }\textbf {\bibinfo
  {volume} {94}},\ \bibinfo {pages} {024015} (\bibinfo {year} {2016})},\
  \bibinfo {note} {doi:10.1103/PhysRevD.94.024015 [arXiv:1512.08707
  [gr-qc]]}\BibitemShut {NoStop}%
\bibitem [{\citenamefont {Beig}\ and\ \citenamefont
  {Schmidt}(2000)}]{Schmidt:2000rta}%
  \BibitemOpen
  \bibfield  {author} {\bibinfo {author} {\bibfnamefont {R.}~\bibnamefont
  {Beig}}\ and\ \bibinfo {author} {\bibfnamefont {B.}~\bibnamefont {Schmidt}},\
  }\href {\doibase doi:10.1007/3-540-46580-4} {\bibfield  {journal} {\bibinfo
  {journal} {Lect. Notes Phys.}\ }\textbf {\bibinfo {volume} {540}},\ \bibinfo
  {pages} {325} (\bibinfo {year} {2000})},\ \bibinfo {note}
  {doi:10.1007/3-540-46580-4 p342, Sec. 3}\BibitemShut {NoStop}%
\bibitem [{\citenamefont {Semer{\'a}k}(2002)}]{Semerak:2002uc}%
  \BibitemOpen
  \bibfield  {author} {\bibinfo {author} {\bibfnamefont {O.}~\bibnamefont
  {Semer{\'a}k}},\ }\href {\doibase doi:10.1142/9789812776938_0004} {\bibfield
  {journal} {\bibinfo  {journal} {Gravitation: following the Prague
  inspiration}\ ,\ \bibinfo {pages} {111}} (\bibinfo {year} {2002})},\ \bibinfo
  {note} {doi:10.1142/9789812776938\_0004 gr-qc/0204025}\BibitemShut {NoStop}%
\bibitem [{\citenamefont {Mars}(2014)}]{Mars:2014wbd}%
  \BibitemOpen
  \bibfield  {author} {\bibinfo {author} {\bibfnamefont {M.}~\bibnamefont
  {Mars}},\ }\href {\doibase doi:10.1007/978-3-319-06349-2_8} {\bibfield
  {journal} {\bibinfo  {journal} {Fundam. Theor. Phys.}\ }\textbf {\bibinfo
  {volume} {177}},\ \bibinfo {pages} {191} (\bibinfo {year} {2014})},\ \bibinfo
  {note} {doi:10.1007/978-3-319-06349-2\_8}\BibitemShut {NoStop}%
\bibitem [{\citenamefont {Pradhan}(2016)}]{Pradhan:2016}%
  \BibitemOpen
  \bibfield  {author} {\bibinfo {author} {\bibfnamefont {P.}~\bibnamefont
  {Pradhan}},\ }\href {\doibase doi:10.1007/s12043-016-1214-x} {\bibfield
  {journal} {\bibinfo  {journal} {Pramana}\ }\textbf {\bibinfo {volume} {87}},\
  \bibinfo {pages} {1} (\bibinfo {year} {2016})},\ \bibinfo {note}
  {doi:10.1007/s12043-016-1214-x [arXiv:1205.5656 [gr-qc]]}\BibitemShut
  {NoStop}%
\bibitem [{\citenamefont {Ono}\ \emph {et~al.}(2016)\citenamefont {Ono},
  \citenamefont {Suzuki},\ and\ \citenamefont {Asada}}]{Ono:2016lql}%
  \BibitemOpen
  \bibfield  {author} {\bibinfo {author} {\bibfnamefont {T.}~\bibnamefont
  {Ono}}, \bibinfo {author} {\bibfnamefont {T.}~\bibnamefont {Suzuki}}, \ and\
  \bibinfo {author} {\bibfnamefont {H.}~\bibnamefont {Asada}},\ }\href
  {\doibase doi:10.1103/PhysRevD.94.064042} {\bibfield  {journal} {\bibinfo
  {journal} {Phys. Rev. D}\ }\textbf {\bibinfo {volume} {94}},\ \bibinfo
  {pages} {064042} (\bibinfo {year} {2016})},\ \bibinfo {note}
  {doi:10.1103/PhysRevD.94.064042 [arXiv:1605.05816 [gr-qc]]}\BibitemShut
  {NoStop}%
\bibitem [{\citenamefont {Izhikevich}(2007)}]{mscodef}%
  \BibitemOpen
  \bibfield  {author} {\bibinfo {author} {\bibfnamefont {E.~M.}\ \bibnamefont
  {Izhikevich}},\ }\href@noop {} {\emph {\bibinfo {title} {Dynamical systems in
  neuroscience}}}\ (\bibinfo  {publisher} {MIT press},\ \bibinfo {year}
  {2007})\BibitemShut {NoStop}%
\bibitem [{\citenamefont {Stephani}\ \emph {et~al.}(2009)\citenamefont
  {Stephani}, \citenamefont {Kramer}, \citenamefont {MacCallum}, \citenamefont
  {Hoenselaers},\ and\ \citenamefont {Herlt}}]{exactsolutionsofes}%
  \BibitemOpen
  \bibfield  {author} {\bibinfo {author} {\bibfnamefont {H.}~\bibnamefont
  {Stephani}}, \bibinfo {author} {\bibfnamefont {D.}~\bibnamefont {Kramer}},
  \bibinfo {author} {\bibfnamefont {M.}~\bibnamefont {MacCallum}}, \bibinfo
  {author} {\bibfnamefont {C.}~\bibnamefont {Hoenselaers}}, \ and\ \bibinfo
  {author} {\bibfnamefont {E.}~\bibnamefont {Herlt}},\ }\href@noop {} {\emph
  {\bibinfo {title} {Exact solutions of Einstein's field equations}}}\
  (\bibinfo  {publisher} {Cambridge University Press},\ \bibinfo {year}
  {2009})\BibitemShut {NoStop}%
\end{thebibliography}%

\end{multicols}
\clearpage
\end{document}